\documentclass[12pt]{article}

\usepackage[dvips]{graphicx}
\renewcommand{\baselinestretch}{1.75}
\font\bigtensl=cmsl10 scaled\magstep2
\newcommand \be{\begin{equation}}
\newcommand \ee{\end{equation}}
\newcommand \ba{\begin{eqnarray}}
\newcommand \ea{\end{eqnarray}}
\begin{document}

\def\today{\ifcase\month\or
 January\or February\or March\or April\or May\or June\or
 July\or August\or September\or October\or November\or December\fi
 \space\number\day, \number\year}
%

\hfil PostScript file created: \today{}; \ time \the\time \ minutes
\vskip .15in

\centerline {LONG- AND SHORT-TERM EARTHQUAKE FORECASTS DURING
THE TOHOKU SEQUENCE
}

\vskip .15in
\begin{center}
{Yan Y. Kagan and David D. Jackson}
\end{center}
\centerline {Department of Earth and Space Sciences,
University of California,}
\centerline {Los Angeles, California 90095-1567, USA;}
\centerline {Emails: {\tt ykagan@ucla.edu,
david.d.jackson@ucla.edu}}
\vskip 0.02 truein

\vspace{0.15in}

\noindent
{\bf Abstract.}
We consider two issues related to the 2011 Tohoku
mega-earthquake:
(1) what is the repeat time for the largest earthquakes in
this area, and
(2) what are the possibilities of numerical short-term
forecasts during the 2011 earthquake sequence in the Tohoku
area.
Starting in 1999 we have carried out long- and short-term
forecasts for Japan and the surrounding areas using the GCMT
catalog.
The forecasts predict the earthquake rate per area, time,
magnitude unit and earthquake focal mechanisms.
Long-term forecasts indicate that the repeat time for the m9
earthquake in the Tohoku area is of the order of 350 years.
We have archived several forecasts made before and after the
Tohoku earthquake.
The long-term rate estimates indicate that, as expected, the
forecasted rate changed only by a few percent after the Tohoku
earthquake, whereas due to the foreshocks, the short-term rate
increased by a factor of more than 100 before the mainshock
event as compared to the long-term rate.
After the Tohoku mega-earthquake the rate increased by a
factor of more than 1000.
These results suggest that an operational earthquake
forecasting strategy needs to be developed to take the
increase of the short-term rates into account.

\vskip .15in
\noindent
{\bf Short running title}:
{\sc
Tohoku earthquake prediction
}

\vskip 0.05in
\noindent
{\bf Key words}:
\vskip .05in
Probabilistic forecasting;
Probability distributions;
Earthquake interaction, forecasting, and prediction;
Statistical seismology;
Time-independent and time-dependent forecasts;
Forecast testing;
Subduction zones;
Maximum/corner magnitude.

\section{Introduction}
\label{intro}

The Tohoku, Japan, magnitude 9.1 earthquake (11 March 2011) and
the ensuing tsunami near the east coast of the island of Honshu
caused nearly 20,000 deaths and more than 300 billion
dollars in damage, resulting in the worst natural disaster
ever recorded (Hayes {\it et al.}, 2011; Simons {\it et al.},
2011; Geller, 2011; Stein {\it et al.}, 2011).

Several quantitative estimates of maximum possible earthquakes
in subduction zones had been published before the Tohoku
event (Kagan, 1997a; Kagan and Jackson, 1994; Bird and
Kagan, 2004; Kagan {\it et al.}, 2010).
In these publications the maximum size of these earthquakes
was determined to be within a range $m 8.5$ -- $m 9.6$.
Two quantitative methods have been deployed to estimate the
maximum size of an earthquake: a statistical determination of
the magnitude--moment/frequency parameters and a moment
conservation principle.
The former technique employs standard statistical parameter
estimation to evaluate two parameters of the earthquake size
distribution: the $b$-value and the maximum magnitude (Kagan,
2002a; 2002b).
The second method works by comparing the estimates of tectonic
deformation at plate boundaries with a similar estimate of
the seismic moment release (Kagan, 1997a).

The statistical estimate of the maximum magnitude for global
earthquakes, including subduction zones and other tectonic
regions, yielded the values $m_{\rm max} \approx 8.3$ (Kagan
and Jackson, 1994; 2000).
The moment conservation provided an estimate for subduction
zones $m_{\rm max} = 8.5$ -- $8.7 \pm 0.3$ (Kagan, 1997a;
1999; 2002b).
The most important result by Kagan (1997a) is that the maximum
earthquake size is the same, at least statistically, for all
the studied subduction zones.
Combined with the estimate of $m_{\rm max} = 9.6$ based on the
analysis of global seismicity (Bird and Kagan, 2004), this
implies that for all major subduction zones the maximum
earthquake magnitude should be greater than 9.0 (Kagan and
Jackson, 2011b).

Our forecasting technique is to establish a statistical model
that fits the catalog of earthquake times, locations, and
seismic moments, and subsequently to base forecasts on this
model.
While most components of the model have been tested
(Kagan and Knopoff, 1987; Kagan, 1991; Jackson and Kagan,
1999; Kagan and Jackson, 2000; Kagan {\it et al.}, 2010), some
require further exploration and can be modified as our
research progresses.

Our previous forecast model was based on constructing a map of
smoothed rates of past earthquakes.
We used the Global Centroid Moment Tensor catalog (Ekstr\"om
{\it et al.}, 2005, referred to subsequently as GCMT)
because it employs relatively consistent methods and lists
tensor focal mechanisms.
The focal mechanisms allow us to estimate the fault plane
orientation for past earthquakes, through which we can
identify a preferred direction for future events.
Using the forecasted tensor focal mechanism, it may be
possible to calculate an ensemble of seismograms for each
point of interest on the Earth's surface.

In Section~\ref{size} we consider two statistical
distributions for the earthquake moment magnitude and show the
magnitude-frequency relations for the Tohoku area.
We evaluate the approximate recurrence interval for a $m
\ge 9.0$ earthquake for the Tohoku area of the order of
350 years (Section~\ref{time}).
Sections~\ref{time} and \ref{short} show the long- and
short-term earthquake forecasts during the Tohoku sequence.
Section~\ref{short} illustrates the possibility of an
operational type calculation of short-term earthquake rates
for short intervals (a few days) after a major earthquake
exceeding the long-term rates by 100--1000 times.

\section{Earthquake size distribution }
\label{size}

We studied earthquake distributions and clustering for the
global CMT catalog of moment tensor inversions compiled by
the GCMT group (Ekstr\"om {\it et al.}, 2005; Ekstr\"om, 2007;
Nettles {\it et al.}, 2011).
The present catalog contains more than 33,000 earthquake
entries for the period 1977/1/1 to 2010/12/31.
The earthquake size is characterized by a scalar seismic
moment $M$.

In analyzing earthquakes here we use the scalar seismic moment
$M$ directly, but for easy comparison and display we convert
it into an approximate moment magnitude using the relationship
(Hanks, 1992)
\be
m_W \ = \ {2 \over 3} \, (\, \log_{10}M - C \, ) \, ,
\label{Eq01}
\ee
where $C = 9.0$, if moment $M$ is measured in Newton m (Nm),
and $C = 16.0$ for moment $M$ expressed in {\sl dyne-cm} as in
the GCMT catalog.
Since we are using the moment magnitude almost exclusively,
later we omit the subscript in $m_W$.
Unless specifically indicated, we use the moment magnitude
calculated as in (\ref{Eq01}) with the scalar seismic moment
from the GCMT catalog.

The earthquake size distribution is usually described by the
G-R (Gutenberg and Richter, 1954) magnitude-frequency
relation
\begin{equation}
{\rm lg} N (m) = a - b \, m \, ,
\label{mom2}
\end{equation}
where $N(m)$ is the number of earthquakes with magnitude $\ge
m$, and $a$ and $b$ are parameters: $a$ characterizes seismic
activity or earthquake productivity of a region and $b$
describes the relation between small and large earthquake
numbers, $b \approx 1$.

The tapered G-R (TGR) distribution includes an exponential
roll-off of frequency for moments near and above a value
called the corner moment (Kagan, 2002a).
The taper ensures that the total moment rate is finite, and
the value of the total moment rate depends strongly on the
corner magnitude.
For magnitudes, the tapered G-R relation can be written as
\be
\log_{10} \, N (m) \ = \ \log_{10} \, a \, - \, b \, (m
- m_t) \, + \, {1 \over {\log \, (10)}} \left [ 10^{1.5 \,
(m_t - m_c)} - 10^{1.5 \, (m - m_c)} \right ]
\, ,
\label{mom3}
\ee
where $m_t$ is the threshold magnitude: the smallest magnitude
above which the catalogue can be considered to be complete,
$m_c$ being s the corner magnitude.
For the standard two-parameter G-R distribution (\ref{mom2}),
the last two terms in the right-hand part of (\ref{mom3}) are
zero ($m_c \to \infty$).

In Fig.~\ref{fig01} we show the moment-frequency curves for
shallow earthquakes (the depth is less or equal to 70~km) in
the Tohoku area (35-40$^\circ$~N, 140-146$^\circ$~E).
The largest earthquake during this period was $m7.68$, thus a
low statistical value for the corner magnitude
($m^s_c=7.8$) is needed to approximate the distribution,
whereas the moment conservation principle yielded the value
$m^m_c=9.3$.

Fig.~\ref{fig02} demonstrates why the values of the maximum
magnitude determined by historical accounts and even by
the standard statistical evaluation method are often grossly
biased downward, especially for small time-space intervals,
as we see in Fig.~\ref{fig01}.
After the Tohoku mega-earthquake ($m9.15$) the statistical
estimate of the corner magnitude changed drastically
($m^s_c=10.97$).
The lower 95\% confidence limit estimate of
the corner magnitude is $m8.9$.
At the same time, the moment conservation value practically
remained the same, i.e., $m^m_c=9.3$.

\section{Long-term earthquake forecasts during the
Tohoku sequence }
\label{time}

Since 1977 we developed statistical models of seismicity
which fit a catalog of earthquake times, locations, and
seismic moments; and subsequently we base our forecasts on
these models.
The forecasts are produced in two formats: long- and
short-term (Kagan and Knopoff, 1977; Kagan and Jackson, 1994;
Jackson and Kagan, 1999; Kagan and Jackson, 2000; 2011a) and
presently they predict the earthquake rate per area, time,
magnitude unit, and focal mechanism.
Several earthquake catalogs are used in our forecasts, the
GCMT catalog was used most frequently as it employs relatively
consistent methods and reports tensor focal mechanisms.
The forecasts including those for the north-west Pacific area
covering Japan, are posted on our Web site:
http://eq.ess.ucla.edu/$\sim$kagan/predictions\_index.html .

Figs.~\ref{fig10} and \ref{fig11} show long-term forecasts
for the north-west Pacific area; one forecast is calculated
before the 2011 Tohoku sequence started, the other one week
after the mega-earthquake.
These forecasts are calculated around midnight Los Angeles
time.
Their description can be found in our publications (Jackson
and Kagan, 1999; Kagan and Jackson, 1994; 2000; 2011).
There is little difference between these forecasts~--
long-term forecasts do not depend strongly on current events.
Plate~1 in Kagan and Jackson (1994) and Fig.~8a in
Kagan and Jackson (2000) display previous forecasts for the
same region.
Appearance of both plots is similar to Figs.~\ref{fig10} and
\ref{fig11}.

In Table~\ref{Table2}, we display earthquake forecast rates
around the epicenter of the $m7.4$ Tohoku foreshock.
The ratio of the short- to long-term rates (the last column)
rises sharply both after the foreshock and after the
mainshock.
Conclusions similar to those in the previous paragraph can be
drawn from Table~\ref{Table2}: the maximum long-term rates
change only by a few tens of a percent.
The predicted focal mechanisms are also essentially the same
for the center of the focal area.

To calculate the earthquake long-term rates for the extended
area we can integrate the tables over the desired surface, or
as a better option, calculate an ensemble of the seismograms
for each point of interest on the Earth's surface.
This can be accomplished by using a forecasted tensor
focal mechanism, such as shown in Table~\ref{Table2}.
These seismograms can be used to calculate probable damage to
any structure due to earthquake waves.
Such calculations are superior in usefulness to earthquake
hazard maps (such as the Japanese map shown, for example, by
Geller, 2011 and by Stein {\it et al.}, 2011).
Hazard maps display an intensity of shaking estimated for
one particular wave period, whereas synthetic seismograms
allow calculating a probability of any structural damage or
collapse, depending on the structure's mechanical properties.

Another advantage of earthquake rate forecasts is that they are
easily tested for effectiveness (Kagan and Jackson, 2000; 2011)
by comparing their predictions with future earthquakes.
This testing information is readily available from earthquake
catalogs.
Hazard maps are more difficult to verify; data on the ground
motion intensity are more scarce, especially in the less
populated territories.
Moreover, these maps may fail for two reasons: an incorrect
seismicity model or an incorrect attenuation relation, thus it
is difficult to find out the cause of poor performance.

As an illustration, using simple methods we make an
approximate estimate of the long-term recurrence rate for
large earthquakes in the Tohoku area.
In the GCMT catalog, the number of earthquakes with $M \ge 5.8$
in a spherical rectangle 35-40$^\circ$~N, 140-146$^\circ$~E,
covering the rupture area of the Tohoku event, is 109 for
years 1977-2010 (Fig.~\ref{fig01}).
If we assume that the corner magnitude is well above $m9.0$
(similar to $m9.6$ for subduction zones, see Bird and Kagan,
2004), the repeat time for the $m9$ and larger events in this
rectangle depends on the assumed $b$-value and is between 300
and 370 years.

Uchida and Matsuzawa (2011) suggest a recurrence interval
for $m9$ events 260-880 years.
Simons {\it et al.}\ (2011) propose a 500-1000 year interval.
This interval estimate is partly based on the observation of
the Jogan earthquake of 13 July 869 and its tsunami.
Even if we assume that this Jogan event was similar in
magnitude to the 2011 earthquake and no other such earthquakes
have occurred meanwhile (see more discussion in Koketsu and
Yokota, 2011), for the Poisson occurrence the probability of
such an interval is of the order 3-5\%.
Moreover, the observation of only one inter-event interval
does not constrain the recurrence time of these
mega-earthquakes in a really meaningful way; the interval can
be as small as a few hundred years or as large as tens of
thousands of years.
Only the moment conservation principle and the tapered
G-R distribution (TGR) distribution provide a reasonable
estimate for this interval.

\section{Short-term forecasts }
\label{short}

Figs.~\ref{fig12}--\ref{fig15} display short-term forecasts
for the north-west Pacific area produced during the initial
period of the Tohoku sequence.
Fig.~\ref{fig12}, calculated before $m7.4$ foreshock, shows a
few weakly `red' spots in those places where earthquakes
occurred during the previous weeks and months.
The short-term rate in these spots is usually of the order of
a few percent or a few tens of a percent compared to the
long-term rate (see also Table~\ref{Table2}).

The predicted earthquake rates in the neighbourhood of the
future Tohoku event increased strongly with the occurrence of
a $m7.4$ foreshock (Fig.~\ref{fig13}).
As Table~\ref{Table2} demonstrate, just before the Tohoku
earthquake, the forecasted rate was about 100 times higher than
the long-term rate.

The area of significantly increased probability covers the
northern part of the Honshu Island following the Tohoku
mega-earthquake occurrence (Fig.~\ref{fig14}).
The size of the area hardly decreased one week later
(Fig.~\ref{fig15}) mostly due to the Tohoku aftershocks.

Although only around 5-10\% few shallow earthquakes are
preceded by foreshocks,
the results shown in Figs.~\ref{fig12}--\ref{fig15} suggest
that an operational earthquake forecasting strategy needs to
be developed (Jordan and Jones, 2010; van Stiphout {\it et
al.}, 2010; Jordan {\it et al.}, 2011) to take the increase of
short-term rates into account.

\section{Discussion }
\label{disc}

It is commonly believed that after a large earthquake the
focal area of an earthquake ``has been destressed" (see, for
example, Matthews {\it et al.}, 2002) thus lowering the
probability of a new large event in this place, though it can
increase in nearby zones.
This reasoning goes back to the flawed seismic
gap/characteristic earthquake model (Jackson and Kagan, 2011).
Kagan and Jackson (1999) showed that earthquakes as large
as 7.5 and larger often occur in practically the same area
soon after the occurrence of a previous earthquake.
Table~\ref{Table3} displays pairs of shallow earthquakes $m
\ge 7.5$ epicentroid of which are closer than their focal zone
size (an update of Table~1 by Kagan and Jackson, 1999, or
Table~1 by Kagan, 2011).
The Table includes three earthquake pairs (see \# 3, 24, 25)
of the Tohoku sequence and demonstrates that strong shocks
tend to repeat in the focal zones of previous events.
Michael (2011) shows that earthquakes as large as $m8.5$ are
clustered in time and space, thus an occurrence of such a big
event does not protect its focal area from the giant next
shock.

Stein {\it et al.}\ (2011) suggest the following forecast
requirements: ideally a forecast should anticipate total
economic and casualty losses due to earthquakes.
Over a relatively short time period, earthquake damage would
seemingly reach the maximum not for the rare very large events,
but for the $m7-m8$ shocks (England and Jackson, 2011).
But over long-term, expected or average losses would peak for
the largest $m8-m9$ events; though their rates are low, the
total average damage for one event increases faster than the
probability of these earthquakes decreases.
Since the expected economic and other losses peak for the
strongest earthquakes (Kagan, 1997b), it is more important to
predict disastrous earthquakes than small ones.
However, the loss calculations (Molchan and Kagan, 1992; Kagan,
1997b) are very uncertain, because major losses are often
caused by unexpected secondary earthquake effects.
Therefore, a prediction of the largest earthquakes is
important, hence prediction schemes that do not specify
the earthquake size are of restricted practical use.
However, if the maximum earthquake is over-predicted, it
diverts resources unnecessarily (Stein {\it et al.}, 2011).

Any forecast scheme that extrapolates the past instrumental
seismicity record would predict future moderate earthquakes
reasonably well.
However as the history of the Tohoku area shows, we need
a different tool to forecast the largest events.
In our forecasts we consider the earthquake rate to be
independent of the earthquake size distribution, so the latter
needs to be specified separately.

As indicated earlier, the seismic moment conservation
principle can provide an answer to the above questions.
The general idea of the moment conservation was suggested
some time ago (Brune, 1968; Wyss, 1973).
However, without the knowledge of the earthquake size
distribution, the calculation of the maximum earthquake moment
size ($m_{\rm max}$) is still difficult and leads to uncertain
or contradictory results.
The classical G-R relation is not helpful in this respect
because it lacks the specification of $m_{\rm max}$.
Only a modification of the G-R law, that introduces
the limiting upper moment could provide a tool to
quantitatively derive $m_{\rm max}$ or its variants.
Kagan and Jackson (2000) and Kagan (2002a, 2002b) propose such
distributions defined by two parameters, $\beta$ and variants
of $m_{\rm max}$.

The application of these distributions allows us also to solve
the problem of evaluating the recurrence period for these
large earthquakes.
Determining the maximum earthquake size either by
historical/instrumental observations or by the qualitative
analogies does not provide such an estimate: a similar
earthquake may occur hundreds or tens of thousand years later.
Fig.~\ref{fig01} shows how using statistical distributions may
facilitate such calculations.

As we discussed in Kagan and Jackson (2011b), the moment
conservation principle allows to quantitatively determine the
maximum earthquake size.
In this respect area-specific calculations provide a more
precise size evaluation for many tectonic zones and, most
importantly to show that the subduction zones could have the
same maximum earthquake size (Kagan, 1997a).
Although the determination of $m_{\rm max}$ by comparing
tectonic and seismic rates is not yet sufficiently accurate
for our purposes, giving $m_{\rm max}$ in the range of 8.5 to
9.7, comparing these estimates to the number of largest
earthquakes in the subduction zones during the last 110 years
definitely argues for the larger of the above values.

In conclusion, we would like to determine the upper magnitude
limit for the subduction zones as well as recurrence intervals
for such earthquakes.
For the tapered G-R (TGR) distribution Bird and Kagan (2004,
Table~5) determined that $m_{cm} = 9.58^{+\infty}_{-0.23}$,
and the 95\% upper limit $m_{cm} = 10.1$.
For the sake of simplicity we take $m_{\rm max} = 10.0$.
Calculations similar to Eq.~8 by Kagan and Jackson (2011b) can
be made to obtain an approximate estimate of the average
inter-earthquake period.
The return period can be estimated from Fig.~1b by Kagan
(2002a) or Eq.~\ref{mom3} as it differs from the regular G-R
law: for the TGR distribution cumulative function at $m_c$ is
below the G-R line by a factor of $e$.
Thus, for the TGR distribution, the recurrence time for the
global occurrence of the $m \ge 10.0$ earthquake is about
475~years.
Of course, the distributions in these calculations are
extrapolated beyond the limit of their parameters' evaluation
range, but the above recurrence periods provide a rough
idea how big such earthquakes can be and how frequently they
can occur worldwide.

For Flinn-Engdahl (Flinn {\it et al.}, 1974) \#19 zone, which
includes Japan, the $m \ge 10.0$ earthquake could repeat in
about 9,000 years for the TGR distribution.
The rupture length of the $m10.0$ event can be estimated from
Fig.~9 by Kagan and Jackson (2011b): at about 2100~km it
is comparable to the 3000~km length of zone \#19.
These long recurrence periods indicate that it would be
difficult to find displacement traces for these earthquakes in
paleo-seismic investigations.

\section{Conclusions }
\label{conc}
{\phantom {\bigtensl {WEB Home Pages: }}}$\,$
\vskip - 1.0 cm

$\bullet$
1.
The major cause for excessive fatalities and economic losses
during the worst global natural disaster in the Tohoku-Oki
area was a gross under-estimation of the maximum earthquake
magnitude ($m_{\rm max}$) and its recurrence interval.

$\bullet$
2.
Long-term forecasts based on the optimal smoothing of
seismicity in and around Japan suggest that the recurrence
period for the $m9$ earthquakes is of the order of 350~years
in the Tohoku area.

$\bullet$
3.
Short-term forecasts can provide time-dependent information
for aftershocks occurrence.
In some cases, if foreshocks are present, as in the Tohoku
sequence, mainshock rates can be predicted.
Therefore, these forecasts can be used for developing an
operational earthquake forecasting strategy.

\subsection* {Acknowledgments
}
\label{Ackn}

We are grateful to Peter Bird and Paul Davis for useful
discussion and suggestions.
The authors appreciate support from the National Science
Foundation through grants EAR-0711515 and EAR-0944218, as well
as from the Southern California Earthquake Center (SCEC).
SCEC is funded by NSF Cooperative Agreement
EAR-0106924 and USGS Cooperative Agreement 02HQAG0008.
Publication 0000, SCEC.

\pagebreak

\def\reference{\hangindent=22pt\hangafter=1}

\pagebreak

\centerline { {\sc References} }
\vskip 0.1in
\parskip 1pt
\parindent=1mm
\def\reference{\hangindent=22pt\hangafter=1}

\reference
Bird, P., and Y. Y. Kagan, 2004.
Plate-tectonic analysis of shallow seismicity: apparent
boundary width, beta, corner magnitude, coupled
lithosphere thickness, and coupling in seven tectonic settings,
{\sl Bull.\ Seismol.\ Soc.\ Amer.}, {\bf 94}(6), 2380-2399,
(plus electronic supplement),
see also an update at
\vspace {-0.25 truecm}
\begin{small}\begin{verbatim}
    http://peterbird.name/publications/2004_global_coupling/2004_global_coupling.htm
\end{verbatim}\end{small}
\vspace {-0.25 truecm}

\reference
Brune, J.~N., 1968.
Seismic moment, seismicity, and rate of slip along major fault
zones,
{\sl J.\ Geophys.\ Res.}, {\bf 73}, 777-784.

\reference
Ekstr\"om, G., 2007.
Global seismicity: results from systematic waveform analyses,
1976-2005,
in {\sl Treatise on Geophysics}, {\bf 4}(4.16), ed.\ H.
Kanamori, pp.~473-481, Elsevier, Amsterdam.

\reference
Ekstr\"om, G., A. M. Dziewonski, N. N. Maternovskaya
and M. Nettles, 2005.
Global seismicity of 2003: Centroid-moment-tensor solutions
for 1087 earthquakes,
{\sl Phys.\ Earth planet.\ Inter.}, {\bf 148}(2-4), 327-351.

\reference
England, P., and J. Jackson, 2011.
Uncharted seismic risk,
{\sl Nature Geoscience}, {\bf 4}, 348-349.

\reference
Flinn, E.\ A., E.\ R.\ Engdahl, and A.\ R.\ Hill, 1974.
Seismic and geographical regionalization,
{\sl Bull.\ Seismol.\ Soc.\ Amer.}, {\bf 64}, 771-992.

\reference
Geller, R. J., 2011.
Shake-up time for Japanese seismology,
{\sl Nature}, {\bf 472}(7344), 407-409,
DOI: doi:10.1038/nature10105.

\reference
Gutenberg, B., and C.\ F.\ Richter, 1954.
{\sl Seismicity of the Earth and Associated Phenomena},
Princeton, Princeton Univ.\ Press., 310~pp.

\reference
Hanks, T.C., 1992.
Small earthquakes, tectonic forces,
{\sl Science}, {\bf 256}, 1430-1432.

\reference
Hayes, G. P., Paul S. Earle, Harley M. Benz, David J. Wald,
Richard W. Briggs the USGS/NEIC Earthquake Response Team,
2011.
88 Hours: The U.S. Geological Survey National Earthquake
Information Center Response to the 11 March 2011 Mw 9.0 Tohoku
Earthquake,
{\sl Seismol.\ Res.\ Lett.}, {\bf 82}(4), 481-493.

\reference
Jackson, D.~D., and Y.~Y.~Kagan, 1999.
Testable earthquake forecasts for 1999,
{\sl Seism.\ Res.\ Lett.}, {\bf 70}(4), 393-403.

\reference
Jackson, D. D., and Y. Y. Kagan, 2011.
Characteristic earthquakes and seismic gaps,
In {\sl Encyclopedia of Solid Earth Geophysics},
Gupta, H. K. (Ed.), Springer, pp.~37-40,
DOI 10.1007/978-90-481-8702-7.

\reference
Jordan, T. H., and L. M. Jones, 2010.
Operational earthquake forecasting: some thoughts on why and
how,
{\sl Seismol.\ Res.\ Lett.}, {\bf 81}(4), 571-574.

\reference
Jordan, T. H., Y.-T. Chen, P. Gasparini, R. Madariaga, I.
Main, W. Marzocchi, G. Papadopoulos, G. Sobolev, K. Yamaoka,
and J. Zschau, 2011.
Operational earthquake forecasting -- state of knowledge and
guidelines for utilization,
{\sl Annals Geophysics}, {\bf 54}(4), 315-391,
doi: 10.4401/ag-5350.

\reference
Kagan, Y.~Y., 1997a.
Seismic moment-frequency relation for shallow earthquakes:
Regional comparison,
{\sl J. Geophys.\ Res.}, {\bf 102}(B2), 2835-2852.

Kagan, Y.~Y., 1997b.
Earthquake size distribution and earthquake insurance,
{\sl Communications in Statistics: Stochastic Models},
{\bf 13}(4), 775-797.

\reference
Kagan, Y.~Y., 1999.
Universality of the seismic moment-frequency relation,
{\sl Pure Appl.\ Geoph.}, {\bf 155}(2-4), 537-573.

\reference
Kagan, Y. Y., 2002a.
Seismic moment distribution revisited: I. Statistical results,
{\sl Geophys.\ J. Int.}, {\bf 148}(3), 520-541.

\reference
Kagan, Y. Y., 2002b.
Seismic moment distribution revisited: II. Moment conservation
principle, {\sl Geophys.\ J. Int.}, {\bf 149}(3), 731-754.

\reference
Kagan, Y. Y., 2011.
Random stress and Omori's law,
{\sl Geophys.\ J. Int.}, {\bf 186}(3), 1347-1364, doi:
10.1111/j.1365-246X.2011.05114.x.

\reference
Kagan, Y. Y., P. Bird, and D. D. Jackson, 2010.
Earthquake patterns in diverse tectonic zones of
the globe,
{\sl Pure Appl.\ Geoph.}\ ({\sl The Frank Evison Volume}),
{\bf 167}(6/7), 721-741, doi: 10.1007/s00024-010-0075-3.

\reference
Kagan, Y.~Y., and D.~D.~Jackson, 1994.
Long-term probabilistic forecasting of earthquakes,
{\sl J. Geophys.\ Res.}, {\bf 99}, 13,685-13,700.

\reference
Kagan, Y.~Y. and D.~D.~Jackson, 1999.
Worldwide doublets of large shallow earthquakes,
{\sl Bull.\ Seismol.\ Soc.\ Amer.}, {\bf 89}(5), 1147-1155.

\reference
Kagan, Y. Y., and D. D. Jackson, 2000.
Probabilistic forecasting of earthquakes,
{\sl Geophys.\ J. Int.}, {\bf 143}, 438-453.

\reference
Kagan, Y. Y. and Jackson, D. D., 2011a.
Global earthquake forecasts,
{\sl Geophys.\ J. Int.}, {\bf 184}(2), 759-776,
doi: 10.1111/j.1365-246X.2010.04857.x.

\reference Kagan, Y. Y. and Jackson, D. D., 2011b. Tohoku
Earthquake: a Surprise? Preprint, http://arxiv.org/abs/1112.5217.

\reference
Kagan, Y., and L. Knopoff, 1977.
Earthquake risk prediction as a stochastic process,
{\sl Phys.\ Earth Planet.\ Inter.}, {\bf 14}(2), 97-108,
doi: 10.1016/0031-9201(77)90147-9.

\reference
Koketsu, K., and Y. Yokota, 2011.
Supercycles along the Japan Trench and Foreseeability of the
2011 Tohoku Earthquake,
AGU Fall Meet.\ Abstract U33C-03.

\reference
Matthews, M. V., W. L. Ellsworth, and P. A. Reasenberg, 2002.
A Brownian model for recurrent earthquakes,
{\sl Bull.\ Seismol.\ Soc.\ Amer.}, {\bf 92}, 2233-2250.

\reference
Michael, A. J., 2011.
Random variability explains apparent global clustering of
large earthquakes,
{\sl Geophys. Res. Lett.}, {\bf 38}, L21301,
doi:10.1029/2011GL049443.

\reference
Molchan, G. M., and Y.~Y.~Kagan, 1992.
Earthquake prediction and its optimization,
{\sl J. Geophys.\ Res.}, {\bf 97}(B4), 4823-4838,
doi:10.1029/91JB03095.

\reference
Nettles, M., Ekstr\"om, G., and H. C. Koss, 2011.
Centroid-moment-tensor analysis of the 2011 off the Pacific
coast of Tohoku Earthquake and its larger foreshocks and
aftershocks,
{\sl Earth Planets Space}, {\bf 63}(7), 519-523.

\reference
Simons, M., Minson, S.E., Sladen, A., Ortega, F., Jiang, J.L.,
Owen, S.E., Meng, L.S., Ampuero, J.P., Wei, S.J., Chu, R.S.,
Helmberger, D.V., Kanamori, H., Hetland, E., Moore, A.W.,
Webb, F.H., 2011.
The 2011 magnitude 9.0 Tohoku-Oki earthquake: mosaicking the
megathrust from seconds to centuries,
{\sl Science}, {\bf 332}(6036), 1421-1425 DOI:
10.1126/science.1206731.

\reference
Stein, S., and E. A. Okal, 2011.
The size of the 2011 Tohoku earthquake need not have been a
surprise,
{\sl Eos Trans.\ AGU}, {\bf 92}(27), 227-228.

\reference
Stein, S., Geller, R., and Liu, M., 2011.
Bad assumptions or bad luck: why earthquake
hazard maps need objective testing,
{\sl Seismol.\ Res.\ Lett.}, {\bf 82}(5), 623-626.

\reference
Uchida, N., and T. Matsuzawa, 2011.
Coupling coefficient, hierarchical structure, and earthquake
cycle for the source area of the 2011 off the Pacific coast of
Tohoku earthquake inferred from small repeating earthquake
data,
{\sl Earth Planets Space}, {\bf 63}(7), 675-679.

\reference
van Stiphout, T., S. Wiemer, and W. Marzocchi (2010).
Are short-term evacuations warranted? Case of the 2009
L'Aquila earthquake,
{\sl Geophys.\ Res.\ Lett.}, {\bf 37}, L06306,
doi:10.1029/2009GL042352.

\reference
Wyss, M., 1973.
Towards a physical understanding of the earthquake frequency
distribution,
{\sl Geophys.\ J.\ R.\ Astr.\ Soc.}, {\bf 31}, 341-359.

\clearpage

\pagebreak

\newpage

\renewcommand{\baselinestretch}{1.25}

\setlength{\oddsidemargin}{-0.5cm}
\setlength{\evensidemargin}{-0.5cm}
\setlength{\textwidth}{17.0cm}

\newpage
\setlength{\topmargin}{-2.5cm}

\begin{table}
\caption{
Examples of long- and short-term forecast
during the Tohoku earthquake sequence.
}
\vspace{10pt}
\label{Table2}
\begin{tabular}{rrcrrrrccl}
\hline
& & & & & & & & & \\[-15pt]
\multicolumn{1}{c}{Lati-}&
\multicolumn{1}{c}{Longi-}&
\multicolumn{6}{c}{LONG-TERM FORECAST}&
\multicolumn{2}{c}{SHORT-TERM}
\\[.3ex]
\multicolumn{1}{c}{tude}&
\multicolumn{1}{c}{tude}&
\multicolumn{1}{c}{Probability}&
\multicolumn{5}{c}{Focal mechanism}&
\multicolumn{1}{c}{Probability}&
\multicolumn{1}{c}{Probability}
\\[.3ex]
\multicolumn{2}{c}{}&
\multicolumn{1}{c}{$m \ge 5.8$}&
\multicolumn{2}{c}{$T$-axis}&
\multicolumn{2}{c}{$P$-axis}&
\multicolumn{1}{c}{Rotation}&
\multicolumn{1}{c}{$m \ge 5.8$}&
\multicolumn{1}{c}{ratio}
\\[.3ex]
\multicolumn{2}{c}{}&
\multicolumn{1}{c}{eq/day*km$^2$}&
\multicolumn{1}{c}{Pl}&
\multicolumn{1}{c}{Az}&
\multicolumn{1}{c}{Pl}&
\multicolumn{1}{c}{Az}&
\multicolumn{1}{c}{angle}&
\multicolumn{1}{c}{eq/day*km$^2$}&
\multicolumn{1}{c}{Time-}
\\[.3ex]
\multicolumn{7}{c}{}&
\multicolumn{1}{c}{degree}&
\multicolumn{1}{c}{Time-}&
\multicolumn{1}{c}{dependent/}
\\[.3ex]
\multicolumn{7}{c}{}&
\multicolumn{1}{c}{}&
\multicolumn{1}{c}{dependent}&
\multicolumn{1}{c}{independent}
\\[2pt]
\hline
\multicolumn{9}{c}{March 8}\\
141.0 & 38.5 & 1.64E-08 & 81 & 307 & 8 & 107 & 24.31 & 5.603E-10 & 3.423E-02\\
141.5 & 38.5 & 2.31E-08 & 76 & 327 & 11 & 107 & 29.62 & 8.171E-10 & 3.535E-02\\
142.0 & 38.5 & 1.04E-07 & 76 & 0 & 5 & 112 & 31.68 & 4.414E-09 & 4.234E-02\\
142.5 & 38.5 & 7.60E-08 & 63 & 312 & 25 & 113 & 11.78 & 2.667E-09 & 3.510E-02\\
143.0 & 38.5 & 4.98E-08 & 60 & 303 & 29 & 113 & 10.25 & 1.215E-09 & 2.441E-02\\
143.5 & 38.5 & 3.06E-08 & 61 & 303 & 28 & 113 & 13.13 & 5.691E-10 & 1.861E-02\\
144.0 & 38.5 & 1.77E-08 & 61 & 306 & 28 & 114 & 15.54 & 1.240E-10 & 6.993E-03\\
144.5 & 38.5 & 1.21E-08 & 55 & 292 & 35 & 117 & 27.55 & 3.989E-10 & 3.300E-02\\
145.0 & 38.5 & 2.08E-08 & 7 & 275 & 82 & 62 & 27.84 & 1.238E-09 & 5.953E-02\\
\multicolumn{9}{c}{March 10}\\
141.0 & 38.5 & 1.64E-08 & 81 & 307 & 8 & 107 & 24.34 & 5.880E-10 & 3.583E-02\\
141.5 & 38.5 & 2.32E-08 & 76 & 327 & 11 & 107 & 29.56 & 1.406E-08 & 0.605\\
142.0 & 38.5 & 1.05E-07 & 76 & 360 & 6 & 112 & 31.66 & 2.482E-06 & 23.7\\
142.5 & 38.5 & 8.14E-08 & 63 & 311 & 26 & 113 & 11.82 & 7.895E-06 & 97.0\\
143.0 & 38.5 & 5.87E-08 & 60 & 301 & 30 & 113 & 9.76 & 5.298E-06 & 90.2\\
143.5 & 38.5 & 3.67E-08 & 61 & 303 & 28 & 113 & 11.19 & 1.868E-06 & 50.9\\
144.0 & 38.5 & 1.78E-08 & 61 & 305 & 28 & 114 & 15.47 & 2.314E-07 & 13.0\\
144.5 & 38.5 & 1.21E-08 & 55 & 292 & 35 & 117 & 27.45 & 3.644E-09 & 0.301\\
145.0 & 38.5 & 2.08E-08 & 7 & 275 & 82 & 63 & 27.91 & 1.241E-09 & 5.963E-02\\
\multicolumn{9}{c}{March 11}\\
141.0 & 38.5 & 1.64E-08 & 81 & 307 & 8 & 107 & 24.35 & 4.960E-06 & 303\\
141.5 & 38.5 & 2.33E-08 & 76 & 326 & 11 & 107 & 29.51 & 9.397E-06 & 404\\
142.0 & 38.5 & 1.05E-07 & 75 & 359 & 6 & 112 & 31.64 & 4.938E-05 & 471\\
142.5 & 38.5 & 8.32E-08 & 63 & 310 & 26 & 113 & 11.80 & 4.271E-05 & 514\\
143.0 & 38.5 & 6.32E-08 & 59 & 301 & 30 & 113 & 9.47 & 3.109E-05 & 492\\
143.5 & 38.5 & 4.10E-08 & 61 & 302 & 29 & 113 & 10.75 & 1.978E-05 & 482\\
144.0 & 38.5 & 2.03E-08 & 60 & 304 & 29 & 114 & 14.44 & 9.430E-06 & 466\\
144.5 & 38.5 & 1.33E-08 & 55 & 293 & 35 & 117 & 25.04 & 5.504E-06 & 413\\
145.0 & 38.5 & 2.14E-08 & 9 & 275 & 81 & 74 & 29.44 & 6.765E-06 & 316\\
&&&&&&&&& \\
\hline
\end{tabular}

\bigskip

\hfil\break
\vspace{5pt}
\end{table}

\newpage

\begin{table}
\caption{Pairs of shallow earthquakes $m \ge 7.5$}
\vspace{10pt}
\label{Table3}
\begin{tabular}{rrrrrrrrrrrrr}
\hline
& & & & & & & & & & & & \\[-15pt]
\multicolumn{1}{c}{}&
\multicolumn{4}{c}{First Event}&
\multicolumn{4}{c}{Second Event}&
\multicolumn{3}{c}{Difference}&
\multicolumn{1}{c}{}
\\[2pt]
\cline{2-5}
\cline{6-9}
\cline{10-12}\\[-1.6ex]
\multicolumn{1}{c}{No}&
\multicolumn{1}{c}{Date}&
\multicolumn{2}{c}{Coord.}&
\multicolumn{1}{c}{$m$}&
\multicolumn{1}{c}{Date}&
\multicolumn{2}{c}{Coord.}&
\multicolumn{1}{c}{$m$}&
\multicolumn{1}{c}{$R$}&
\multicolumn{1}{c}{$\Phi$}&
\multicolumn{1}{c}{$\Delta t$}&
\multicolumn{1}{c}{$\eta$}\\[.3ex]
\multicolumn{2}{c}{}&
\multicolumn{1}{c}{Lat.}&
\multicolumn{1}{c}{Long.}&
\multicolumn{2}{c}{}&
\multicolumn{1}{c}{Lat.}&
\multicolumn{1}{c}{Long.}&
\multicolumn{1}{c}{}&
\multicolumn{1}{c}{km}&
\multicolumn{1}{c}{$^\circ$}&
\multicolumn{1}{c}{day}&
\multicolumn{1}{c}{}
\\[2pt]
\hline
& & & & & & & & & & & & \\[-15pt]
1 & 1977/06/22 & -22.9 & -174.9 & 8.1 & 2009/03/19 & -23.1 & -174.2 & 7.7 & 75 & 55 & 11593.26 & 1.4 \\
2 & 1978/03/23 & 44.1 & 149.3 & 7.6 & 1978/03/24 & 44.2 & 149.0 & 7.6 & 25 & 7 & 1.69 & 2.3 \\
3 & 1978/06/12 & 38.0  & 142.1 & 7.7 & 2011/03/11 & 37.5  & 143.1 & 9.2 & 104 & 12 & 11959.90  & 4.4 \\
4 & 1980/07/08 & -12.9 & 166.2 & 7.5 & 1980/07/17 & -12.4 & 165.9 & 7.8 & 62 & 18 & 8.85 & 1.1 \\
5 & 1980/07/08 & -12.9 & 166.2 & 7.5 & 1997/04/21 & -13.2 & 166.2 & 7.8 & 33 & 42 & 6130.53 & 2.0 \\
6 & 1980/07/08 & -12.9 & 166.2 & 7.5 & 2009/10/07 & -12.6 & 166.3 & 7.7 & 37 & 13 & 10682.95 & 1.6 \\
7 & 1980/07/17 & -12.4 & 165.9 & 7.8 & 2009/10/07 & -12.6 & 166.3 & 7.7 & 41 & 14 & 10674.10 & 1.8 \\
8 & 1980/07/17 & -12.4 & 165.9 & 7.8 & 2009/10/07 & -11.9 & 166.0 & 7.9 & 65 & 12 & 10674.11 & 1.3 \\
9 & 1983/03/18 & -4.9 & 153.3 & 7.8 & 2000/11/16 & -4.6 & 152.8 & 8.1 & 83 & 72 & 6452.83 & 1.3 \\
10 & 1983/03/18 & -4.9 & 153.3 & 7.8 & 2000/11/16 & -5.0 & 153.2 & 7.9 & 47 & 91 & 6452.94 & 1.8 \\
11 & 1985/09/19 & 17.9 & -102.0 & 8.0 & 1985/09/21 & 17.6 & -101.4 & 7.6 & 71 & 14 & 1.51 & 1.3 \\
12 & 1987/03/05 & -24.4 & -70.9 & 7.6 & 1995/07/30 & -24.2 & -70.7 & 8.1 & 33 & 7 & 3068.83 & 2.8 \\
13 & 1990/04/18 & 1.3 & 123.3 & 7.7 & 1991/06/20 & 1.0 & 123.2 & 7.6 & 37 & 29 & 427.65 & 1.6 \\
14 & 1995/08/16 & -5.5 & 153.6 & 7.8 & 2000/11/16 & -5.0 & 153.2 & 7.9 & 76 & 74 & 1918.89 & 1.1 \\
15 & 1997/04/21 & -13.2 & 166.2 & 7.8 & 2009/10/07 & -12.6 & 166.3 & 7.7 & 70 & 30 & 4552.42 & 1.0 \\
16 & 2000/06/04 & -4.7 & 101.9 & 7.9 & 2007/09/12 & -3.8 & 101.0 & 8.6 & 150 & 85 & 2655.78 & 1.3 \\
17 & 2000/11/16 & -4.6 & 152.8 & 8.1 & 2000/11/16 & -5.0 & 153.2 & 7.9 & 67 & 89 & 0.12 & 1.7 \\
18 & 2000/11/16 & -4.6 & 152.8 & 8.1 & 2000/11/17 & -5.3 & 152.3 & 7.8 & 93 & 88 & 1.67 & 1.2 \\
19 & 2001/06/23 & -17.3 & -72.7 & 8.5 & 2001/07/07 & -17.5 & -72.4 & 7.7 & 34 & 8 & 13.54 & 4.7 \\
20 & 2005/03/28 & 1.7 & 97.1 & 8.7 & 2010/04/06 & 2.0 & 96.7 & 7.8 & 58 & 7 & 1835.25 & 4.0 \\
21 & 2006/11/15 & 46.7 & 154.3 & 8.4 & 2007/01/13 & 46.2 & 154.8 & 8.2 & 70 & 82 & 58.71 & 2.7 \\
22 & 2007/09/12 & -3.8 & 101.0 & 8.6 & 2007/09/12 & -2.5 & 100.1 & 7.9 & 176 & 11 & 0.53 & 1.2 \\
23 & 2007/09/12 & -3.8 & 101.0 & 8.6 & 2010/10/25 & -3.7 & 99.3 & 7.9 & 189 & 8 & 1139.15 & 1.1 \\
24 & 2011/03/11 & 37.5 & 143.1 & 9.2 & 2011/03/11 & 35.9 & 141.4 & 8.0 & 232 & 7 & 0.020 & 2.1 \\
25 & 2011/03/11 & 37.5 & 143.1 & 9.2 & 2011/03/11 & 38.3 & 144.6 & 7.7 & 162 & 62 & 0.027 & 2.9 \\
\hline
\end{tabular}


\bigskip
{\sl Notes:}
$R$ -- centroid distance,
$\Phi$ -- 3-D rotation angle between focal mechanisms,
$\Delta t$ -- time interval between events,
$\eta$ -- degree of zone overlap, the ratio of
earthquake focal zone sizes to twice their distance, see
Equations (2,3) in Kagan and Jackson (1999).
The total earthquake number with magnitude $m \ge 7.50$
for the period 1976/1/1--2011/09/20 is 126.
The maximum epicentroid distance is 250.00~km.

\hfil\break
\vspace{5pt}
\end{table}

\newpage

\renewcommand{\baselinestretch}{1.25}

\setlength{\oddsidemargin}{0.0cm}
\setlength{\evensidemargin}{0.0cm}
\setlength{\textwidth}{16.5cm}

\clearpage

\newpage

\renewcommand{\baselinestretch}{1.75}

\parindent=0mm

\begin{figure}
\begin{center}
\includegraphics[width=0.65\textwidth]{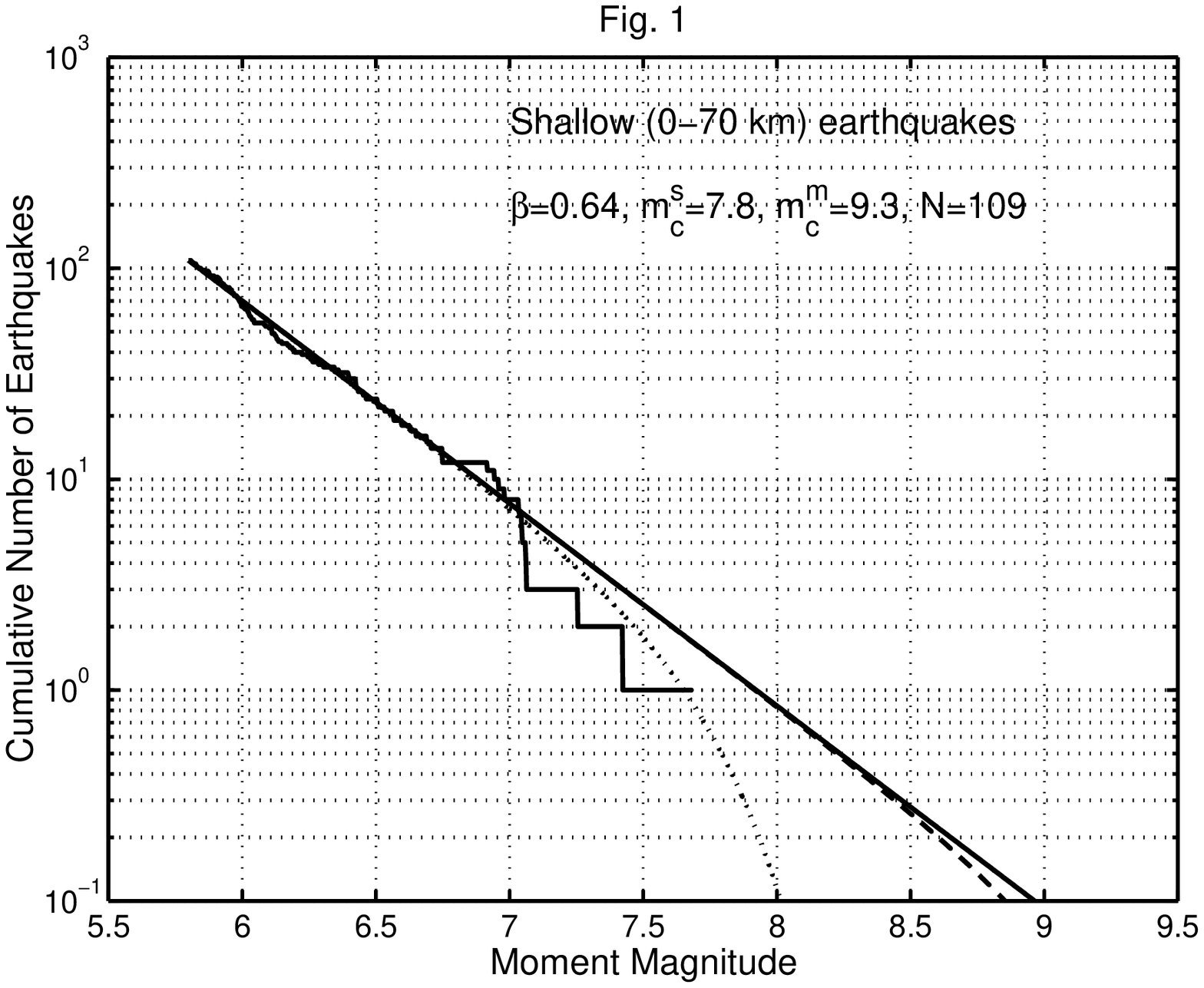}
%
\caption{\label{fig01}
}
\end{center}
The number of earthquakes in the Tohoku area (35-40$^\circ$~N,
140-146$^\circ$~E) with the moment magnitude $(m)$ larger than
or equal to $m$ as a function of $m$ for the shallow
earthquakes in the GCMT catalog during 1977--2010.
Magnitude threshold $m_t=5.8$, the total number of events is
109.
The unrestricted Gutenberg-Richter law is shown by a solid
line.
Dashed and dotted lines show two tapered G-R distributions:
the G-R law restricted at large magnitudes by an exponential
taper with a corner magnitude.
One corner magnitude $m_c^s=7.8$ is evaluated by the maximum
likelihood method using the earthquake statistical record,
another
estimate $m_c^m=9.3$ is based on the moment conservation.
The slope of the linear part of the curve corresponds to
$\beta=0.640$.
\end{figure}

\begin{figure}
\begin{center}
\includegraphics[width=0.65\textwidth]{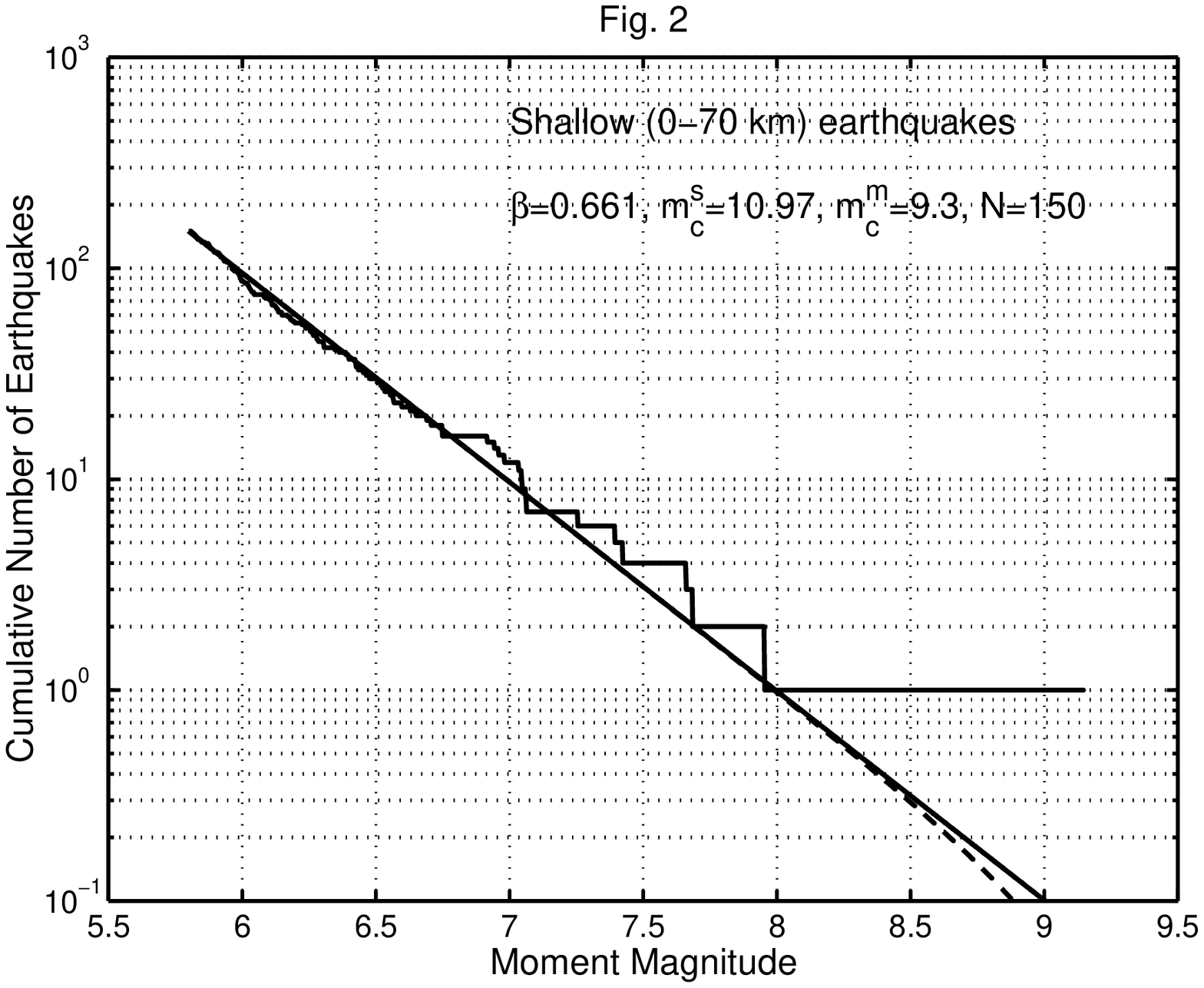}
%
\caption{\label{fig02}
}
\end{center}
Magnitude-frequency relation for the Tohoku area.
The plot is similar to Fig.~\ref{fig01}, but the time interval
is 1977--2011; the total number of events is 150.
One corner magnitude $m_c^s=10.97$ is evaluated by the maximum
likelihood method, another estimate $m_c^m=9.3$ is based on
the moment conservation (Kagan and Jackson, 2011b).
The slope of the linear part of the curve corresponds to
$\beta=0.661$.
Because of the high value of the corner magnitude
($m_c^s=10.97$), one of the curves for the TGR distribution
practically overlays the G-R straight line.
\end{figure}

\begin{figure}
\begin{center}
\includegraphics[width=0.55\textwidth]{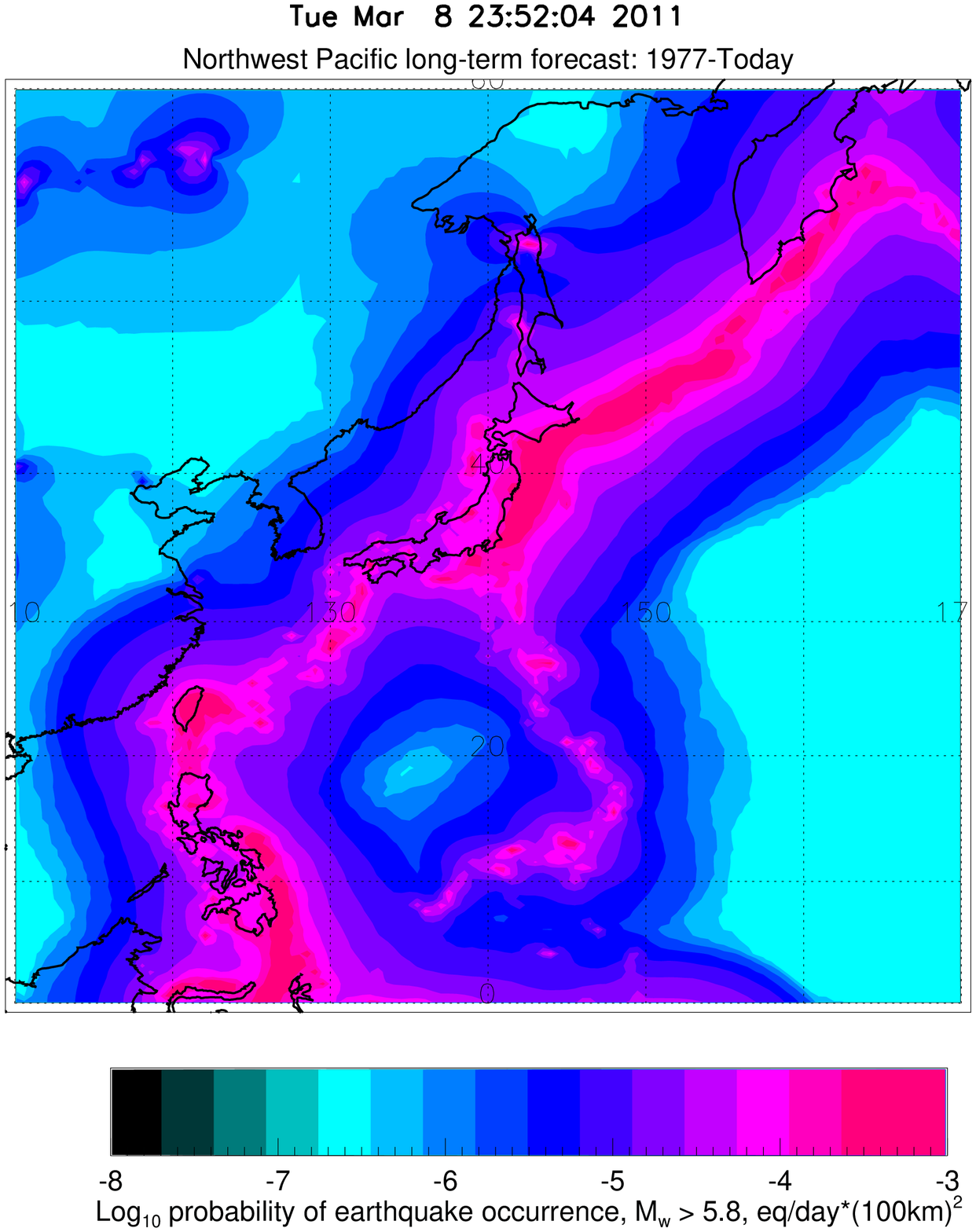}
%
\caption{\label{fig10}
}
\end{center}
The long-term forecast rate for the north-west (NW) Pacific
calculated March 8, 2011, before the $m7.4$ Tohoku foreshock.
\end{figure}

\begin{figure}
\begin{center}
\includegraphics[width=0.55\textwidth]{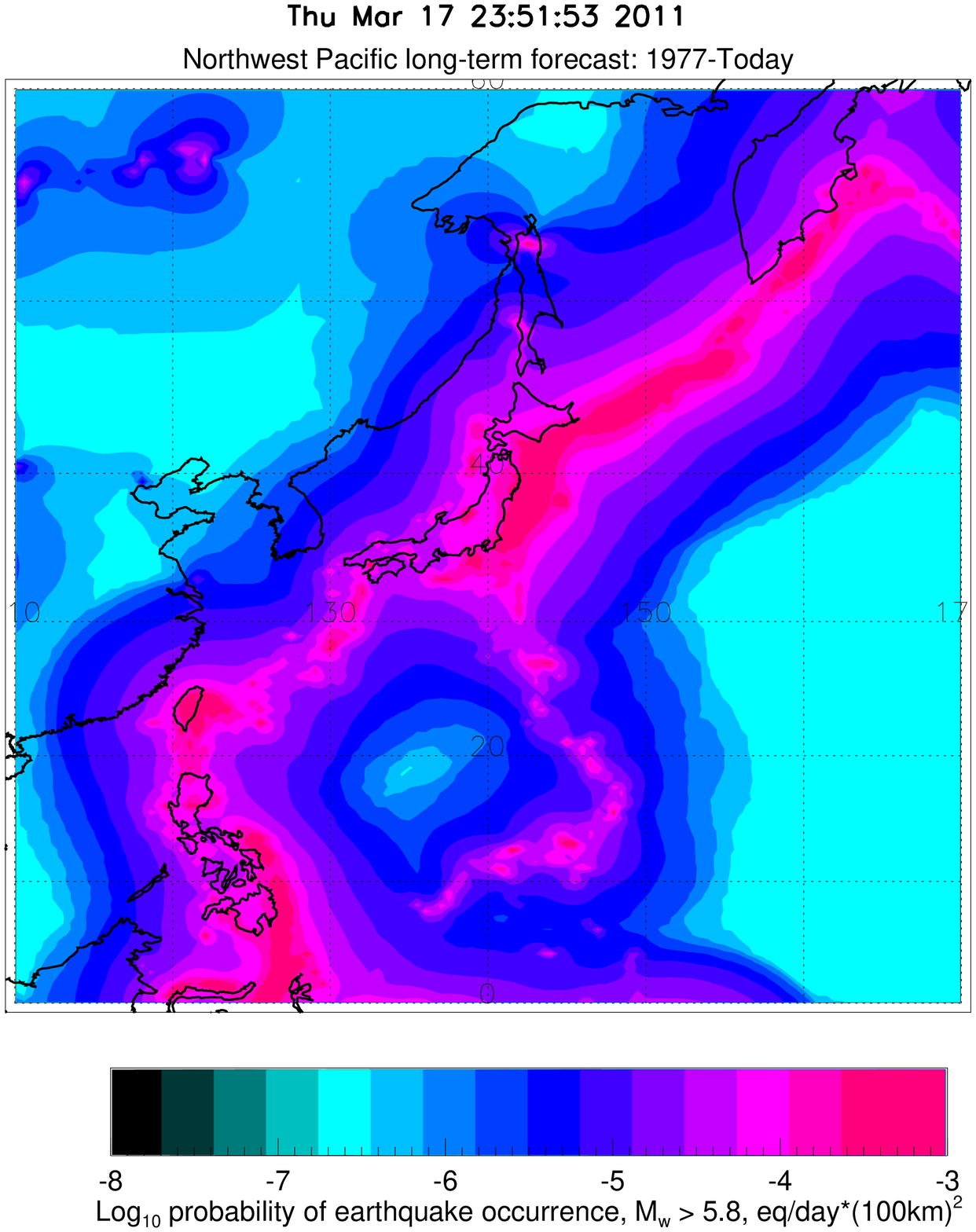}
%
\caption{\label{fig11}
}
\end{center}
The long-term forecast rate for the NW Pacific calculated
March 17, 2011, after the Tohoku mainshock.
There is little change compared to Fig.~\ref{fig10}.
\end{figure}

\begin{figure}
\begin{center}
\includegraphics[width=0.55\textwidth]{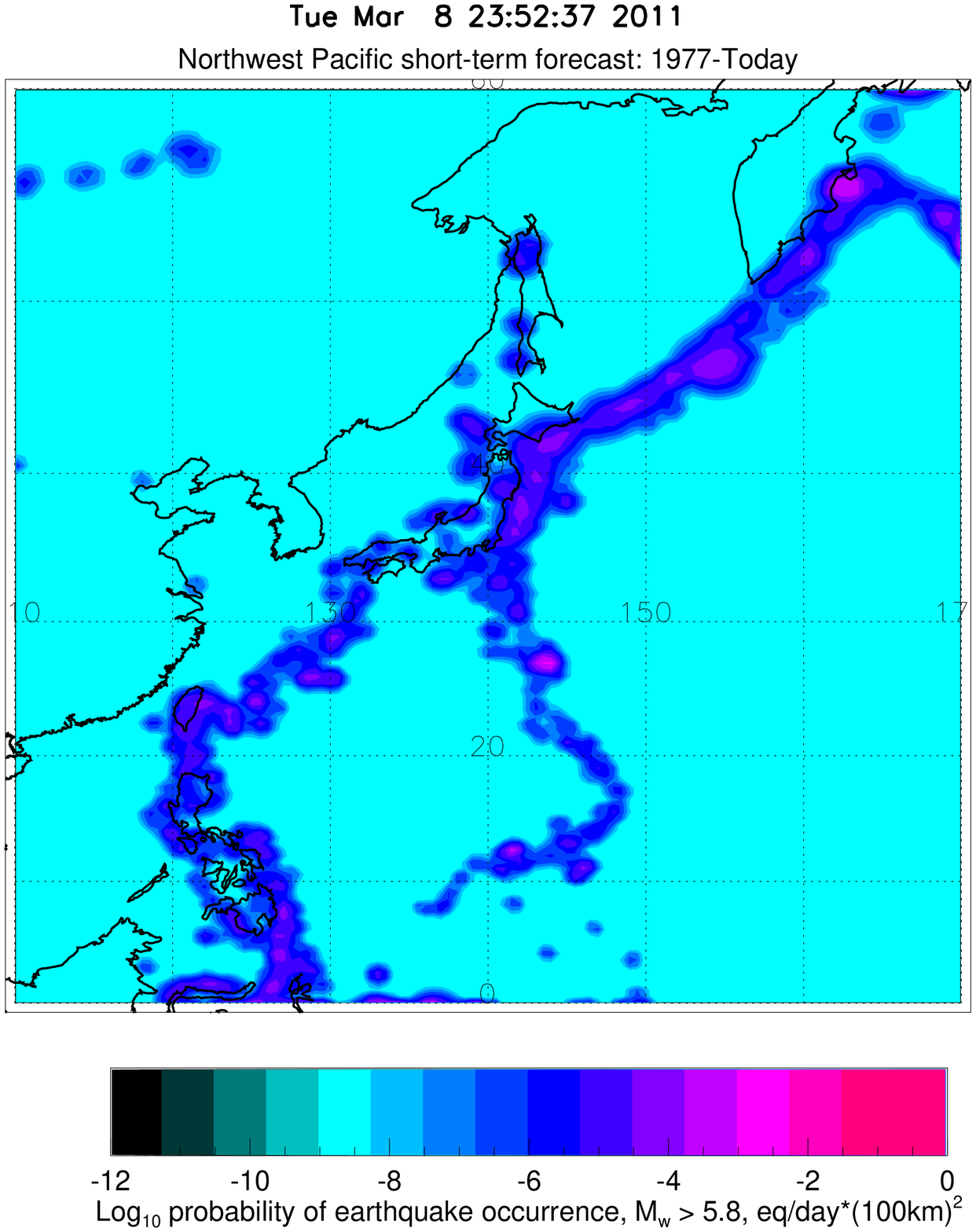}
%
\caption{\label{fig12}
}
\end{center}
The short-term forecast rate for the NW Pacific calculated
March 8, 2011, before the $m7.4$ foreshock.
\end{figure}

\begin{figure}
\begin{center}
\includegraphics[width=0.55\textwidth]{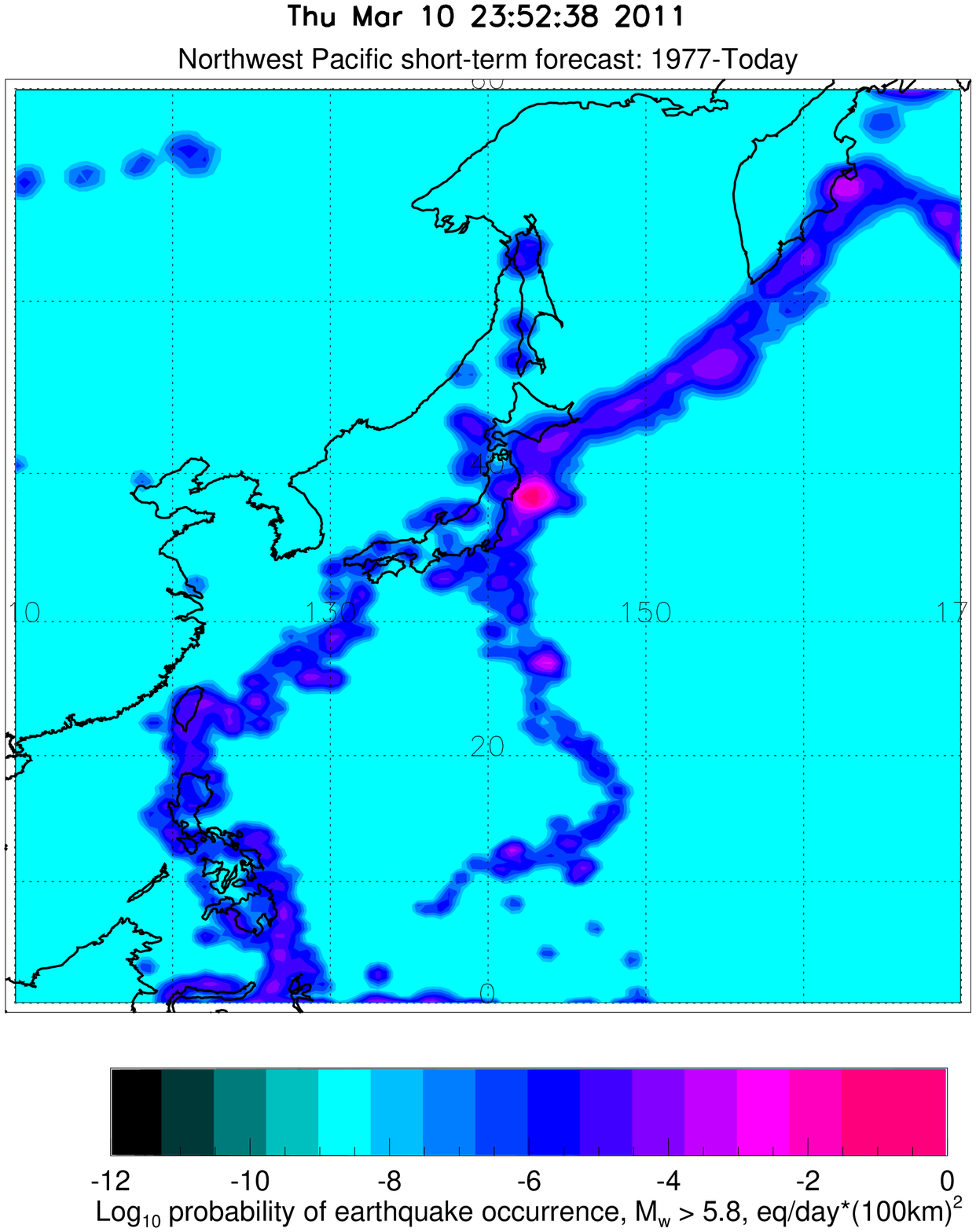}
%
\caption{\label{fig13}
}
\end{center}
The short-term forecast rate for the NW Pacific calculated
March 10, 2011, after the $m7.4$ foreshock, just before the
Tohoku $m9.1$
mainshock -- at the foreshock epicenter the short-term rates
are about 100 times higher than the long-term rates.
\end{figure}

\begin{figure}
\begin{center}
\includegraphics[width=0.55\textwidth]{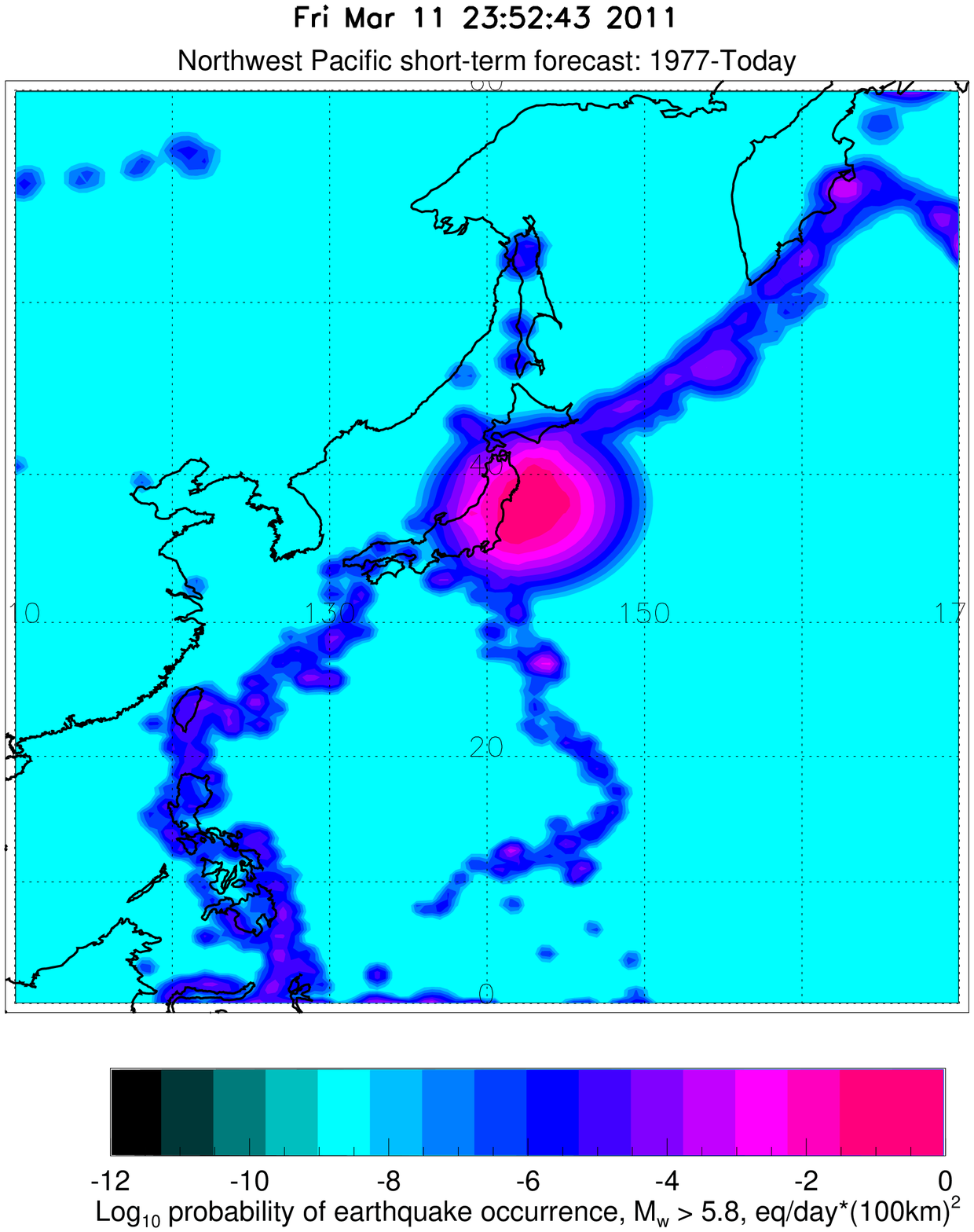}
%
\caption{\label{fig14}
}
\end{center}
The short-term forecast rate for the NW Pacific calculated
March 11, 2011, immediately after the $m9.1$ mainshock,
the short-term rates are about 1000 times higher than the
long-term rates.
\end{figure}

\begin{figure}
\begin{center}
\includegraphics[width=0.55\textwidth]{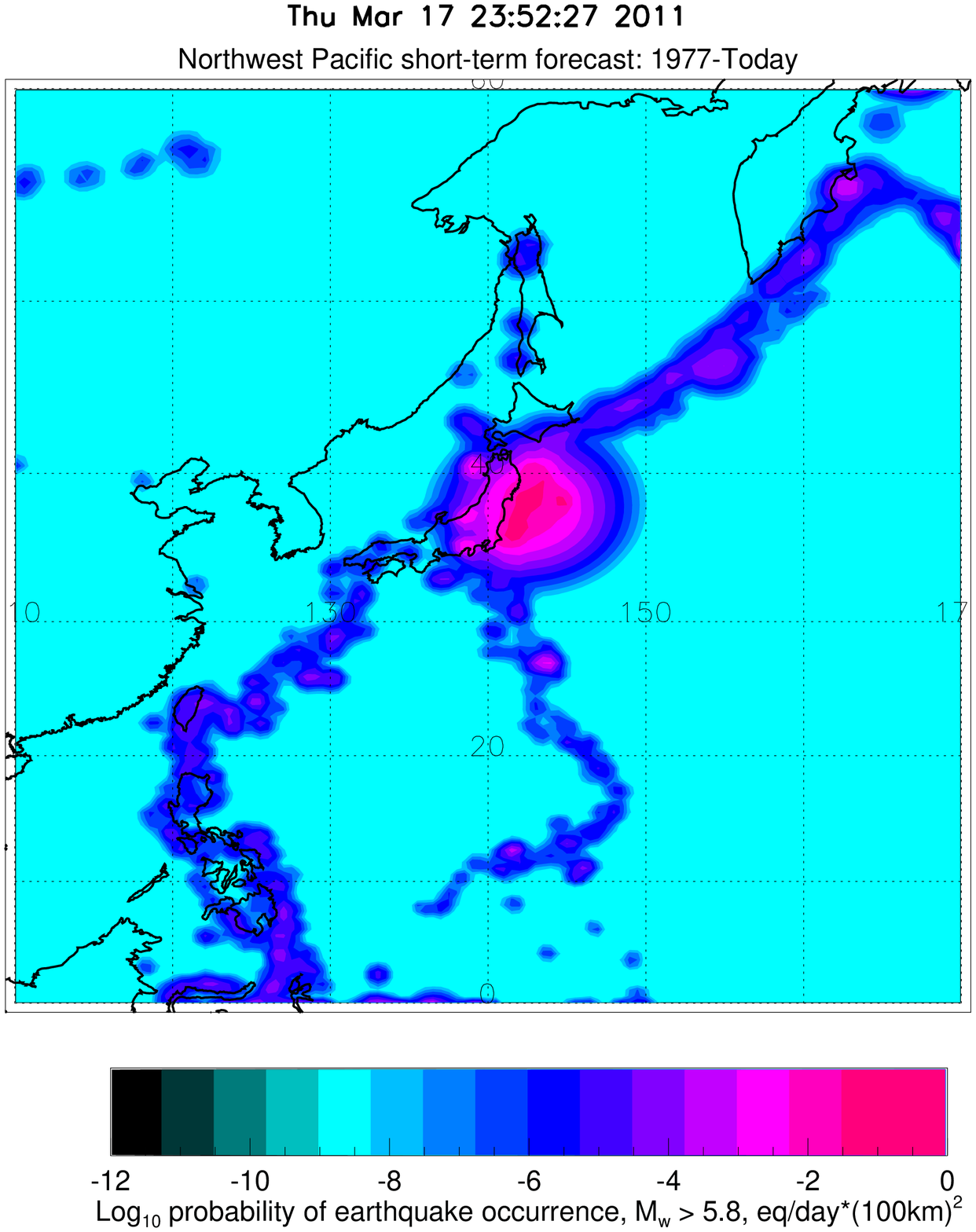}
%
\caption{\label{fig15}
}
\end{center}
The short-term forecast rate for the NW Pacific calculated
March 17, 2011, a week after the $m9.1$ mainshock, the
short-term rates are about 100 times higher than the long-term
rates.
\end{figure}

\newpage

\end{document}